\documentclass[sigconf]{acmart}

\usepackage[acronym]{glossaries}
\usepackage{caption}
\usepackage[most]{tcolorbox}
\usepackage{multirow}
\usepackage{stfloats}
\usepackage{tikz}
\usepackage{tabularx}
\usepackage{xcolor} 
\usepackage{booktabs}
\usepackage{graphicx} 

\tcbuselibrary{breakable}
\tcbset{
    colback=gray!5!white, 
    colframe=gray!80!black, 
    sharp corners,
    breakable,
    enhanced, 
  segmentation style={draw=none},
  bottomrule at break=0pt,
  toprule at break=0pt,
}

\usepackage[utf8]{inputenc}
\usepackage[inline]{enumitem}
    
\tcbuselibrary{breakable}

\AtBeginDocument{%
  }

\begin{document}
\settopmatter{printacmref=false}
\settopmatter{printacmref=false}
\setcopyright{none}
\renewcommand\footnotetextcopyrightpermission[1]{}
\pagestyle{plain}
 \title{RAG-DIVE: A Dynamic Approach for Multi-Turn Dialogue Evaluation in Retrieval-Augmented Generation}
 \thanks{Accepted for publication at CAIN 2026 (5th International Conference on AI Engineering).}
\author{Lorenz Brehme}
\email{lorenz.brehme@uibk.ac.at}
\orcid{0009-0009-4711-2564}
\affiliation{%
  \institution{University of Innsbruck}
  \city{Innsbruck}
  \state{Tirol}
  \country{Austria}
}
\author{Benedikt Dornauer}
\orcid{0000-0002-7713-4686}
\email{benedikt.dornauer@uibk.ac.at}
\affiliation{%
  \institution{University of Innsbruck}
  \city{Innsbruck}
  \state{Tirol}
  \country{Austria}
}

\author{Jan-Henrik Böttcher}
\email{boettcherj@uni-hildesheim.de}
\orcid{0009-0009-4711-2564}
\affiliation{%
  \institution{University of Hildesheim}
  \city{Hildesheim}
  \state{Niedersachsen}
  \country{Germany}
}
\author{Klaus Schmid}
\orcid{0000-0002-7713-4686}
\email{schmid@uni-hildesheim.de}
\affiliation{%
  \institution{University of Hildesheim}
  \city{Hildesheim}
  \state{Niedersachsen}
  \country{Germany}
}

\author{Mircea-Cristian Racasan}
\orcid{0009-0008-7938-3126}
\email{mracasan@cccom.at}
\affiliation{%
  \institution{c.c.com Moser GmbH}
  \streetaddress{Teslastraße 4, 8074}
  \city{Grambach}
  \country{Austria}}

\author{Ruth Breu}
\orcid{0000-0001-7093-4341}
\email{ruth.breu@uibk.ac.at}
\affiliation{%
  \institution{University of Innsbruck}
  \city{Innsbruck}
  \state{Tirol}
  \country{Austria}
}

\renewcommand{\shortauthors}{Brehme et al.}
\acmArticleType{Research}


\definecolor{mygold}{HTML}{E3C800}
\definecolor{mygoldborder}{HTML}{B09500}

\DeclareRobustCommand*\goldcircled[1]{%
  \tikz[baseline=(char.base)]{
    \node[
      shape=circle,
      draw=mygoldborder ,   
      fill=mygold,   
      text=black,   
      thick,
      inner sep=0.0pt,
      minimum size=10pt, 
    ] (char) {#1};}}


\begin{abstract}
Evaluating Retrieval-Augmented Generation (RAG) systems using static multi-turn datasets fails to capture the dynamic nature of real-world dialogues. Existing evaluation methods rely on predefined datasets, which restrict them to static, one-directional queries and limit their ability to capture the adaptive, context-dependent performance of RAG systems in interactive, multi-turn settings.

Thus, we introduce the \href{https://github.com/lorenzbrehme/RAG-DIVE.git}{RAG-DIVE},  a Dynamic Interactive Validation and Evaluation approach, that simulates user interactions with RAG systems. RAG-DIVE leverages an LLM to generate multi-turn conversations dynamically and is organized into three components. The dialogue generation stage consists of the \textbf{(1) Conversation Generator}, which simulates a user by creating multi-turn queries, and the \textbf{(2) Conversation Validator}, which filters and corrects invalid or low-quality outputs to ensure coherent conversations. The evaluation stage is handled by the \textbf{(3) Conversation Evaluator}, which assesses the RAG system’s performance across the entire dialogue and generates both per-turn and multi-turn metrics that provide an aggregated view of system behavior.

We validated RAG-DIVE through two experimental setups.
First, we tested a sample RAG system, including human evaluation of dialogue quality, repeated trials to assess consistency, and an ablation study showing that RAG-DIVE detects performance changes caused by system modifications.
Second, we compared RAG-DIVE with a traditional static dataset evaluation on an industrial RAG system under different configurations to verify whether both approaches reveal similar performance trends.

Our findings demonstrate that RAG-DIVE facilitates dynamic, interaction-driven evaluation for multi-turn conversations, thereby advancing the assessment of RAG systems. 
\end{abstract}

\maketitle

\section{Introduction}
Retrieval-Augmented Generation (RAG) is a rapidly growing area of research that addresses key limitations of large language models (LLMs), such as hallucination and restricted knowledge scope~\cite{lewis_retrieval-augmented_2021}.
In software engineering, it has been applied to several contexts, such as commit message generation~\cite{shi_race_2022} and access control policy extraction ~\cite{aboukadri_leveraging_2025}.
Thereby, the challenge is to realistically and repeatedly evaluate the RAG performance to achieve optimal retrieval and generator performance~\cite{katsis_mtrag_2025, salemi_evaluating_2024}. Parameters such as context size, embedding model, or history implementation can be tuned, and systematic evaluation is necessary to understand how these changes impact the results~\cite{brehme_can_2025}. 

To evaluate the impact of changes, several frameworks --- such as Ragas~\cite{es_ragas_2023} --- have been developed to enable automated evaluation of RAG systems. These frameworks typically assess each RAG component individually. An example query is used as input, and the system’s outputs are evaluated based on multiple criteria. Specifically, the relevance of the retrieved context to the input query is analyzed, and the generated output is assessed in terms of its Correctness and Faithfulness to the retrieved information~\cite{brehme_can_2025}.

To effectively evaluate a RAG system, it is crucial to simulate realistic human–RAG interactions ~\cite{brehme_retrieval-augmented_2025}.
These interactions typically produce a set of question–answer (QA) pairs~\cite{brehme_can_2025}, which can be further extended into multi-turn conversations~\cite{katsis_mtrag_2025}, where a complete conversation is created with several questions and answers that build on each other. QA pairs are widely used for evaluating RAG systems~\cite{brehme_can_2025} and can be generated by either humans or LLMs~\cite{brehme_can_2025, cheng_coral_2024}. During evaluation, the QA sets serve as inputs to the RAG system, where both the retriever and generator outputs are assessed using metrics such as context relevance, faithfulness, and correctness~\cite{cheng_coral_2024, es_ragas_2023}.

The creation of such datasets has been explored in several existing frameworks~\cite{kuo_rad-bench_2024,es_ragas_2023} with several approaches. Some utilize pre-existing QA datasets, which can be directly applied; however, these are often not specifically designed for RAG systems~\cite{brehme_can_2025}. Other methods involve human-curated datasets~\cite{katsis_mtrag_2025} or datasets generated by LLMs~\cite{es_ragas_2023}. Most of these resources consist primarily of single-turn QA pairs, meaning each interaction involves only one question and one answer, and therefore lack support for multi-turn conversations. These QA pairs do not evaluate how the RAG system performs in extended conversations, whether it forgets previously provided information, or how it manages conversation history~\cite{laban_llms_2025}. Nevertheless, recent methods have been proposed to generate multi-turn conversational datasets tailored for RAG evaluation~\cite{lee_multi-document_2024, cheng_coral_2024}. 
However, most existing multi-turn conversation datasets are static ~\cite{levi_intellagent_2025}, follow predetermined sequences, and are not influenced by the RAG system’s actual responses. This setup does not accurately reflect realistic RAG usage. 
Despite the availability of automated RAG evaluation tools and QA datasets, existing methods largely rely on static interactions and fail to model dynamic, multi-turn, and adaptive conversations reflective of real-world use. This limits their effectiveness in evaluating the true capabilities of RAG systems.
To address this limitation, we introduce in Section \ref{Evalaution_strategy} a dynamic and interactive validation and evaluation framework, called RAG-DIVE. RAG-DIVE generates adaptive, multi-turn conversations by interacting directly with the RAG system and evaluates the generated outputs. The framework is validated in Section \ref{valdiation_ragdive} by verifying the realism and reliability of the generated conversations. Furthermore, in Section \ref{industrial_usecase} we additionally adapted RAG-DIVE for an industrial use case  and evaluated its performance against a commonly used evaluation framework based on the static SQuAD dataset to demonstrate its effectiveness.

\section{Related Work}
Evaluating a RAG system requires a dataset that serves as input to the model. Several types of datasets are commonly used. One basic type is a \textit{single-hop} QA dataset, which typically consists of a question, a ground truth answer, and a context passage that is relevant for answering the question~\cite{brehme_can_2025}. Various such datasets exist, and some evaluation frameworks use publicly available resources like the SQuAD dataset~\cite{rajpurkar_know_2018} or the HotPotQA dataset ~\cite{yang2018hotpotqa}. HotPotQA is based on public domain content and includes a variety of question types, including multi-hop questions~\cite{yang2018hotpotqa}. \textit{Multi-hop} queries are those that require reasoning over multiple chunks to arrive at a correct answer~\cite{tang_multihop-rag_2024,trivedi2021musique}. 
Another type of QA dataset is one that is newly created specifically for the RAG system being evaluated, which can include both single-hop and multi-hop questions ~\cite{tang_multihop-rag_2024}. These datasets can be generated either by humans~\cite{schimanski_climretrieve_2024, li_benchmarking_2024} or by LLMs ~\cite{tang_multihop-rag_2024, es_ragas_2023}. In this case, the questions are crafted based on context documents from the RAG system’s own knowledge base, making the dataset more tailored to the specific system~\cite{es_ragas_2023}. 

Building on these QA datasets, recent work has extended the evaluation paradigm from isolated \textit{single-turn} questions to \textit{multi-turn} conversations.
This shift better reflects realistic usage of RAG systems, where users interact over multiple exchanges rather than through one-off queries. In such settings, the system must not only answer correctly but also retain and apply information across turns~\cite{katsis_mtrag_2025}. Corresponding multi-turn datasets have been developed either manually~\cite{katsis_mtrag_2025} or automatically with LLMs~\cite{cheng_coral_2024, sirdeshmukh_multichallenge_2025}, again grounded in the system’s underlying knowledge base.

To increase the realism and depth of such multi-turn conversations, ~\citet{sirdeshmukh_multichallenge_2025} introduced six question types, each targeting different conversational dynamics. \textit{Direct} questions are posed independently, without relying on prior context. \textit{Follow-up} questions build on information from earlier turns, while \textit{comparative} questions require evaluating or contrasting multiple entities. \textit{Clarification} questions seek to resolve ambiguity or request further detail, and \textit{correction} questions address misunderstandings or errors in previous responses. \textit{Unanswerable} questions represent cases where no valid answer can be derived from the provided context.

All of these datasets are typically used as static inputs to the RAG system, which then produces an output of the RAG. The quality of this output must be evaluated, and to facilitate this, various automated evaluation frameworks have been developed. One common approach involves using traditional Natural Language Processing metrics such as BLEU or ROUGE, which measure the lexical similarity between the RAG-generated answer and the ground truth answer provided in the dataset~\cite{katsis_mtrag_2025, cheng_coral_2024}.

Another approach involves using an LLM-based evaluator to automatically assess the performance of the RAG system across specific metrics. In this setup, the input query, the retrieved context, and the generated answer are provided to the evaluator to compute quality scores~\cite{brehme_can_2025}.
One such metric is Correctness, which measures whether the generated answer aligns with the ground truth answer and the information found in the provided context documents~\cite{yu_knowledge-centric_2024, yang_crag_2024}. Another key metric is Faithfulness, where the LLM determines whether the answer is factually consistent with the retrieved documents~\cite{es_ragas_2023}. To evaluate the quality of the retrieval step, the metric context relevance is used, assessing whether each retrieved document is relevant to answering the query~\cite{saad-falcon_ares_2024}. Studies have shown a positive correlation between LLM-based and human evaluation ~\cite{brehme_can_2025}.

Additionally, metrics can be differentiated using single-turn-level metrics, where each turn is evaluated, or multi-turn-level metrics, where the entire dialogue is evaluated ~\cite{acikgoz_td-eval_2025}.
The Ragas ~\cite{es_ragas_2023} framework provides various single-turn metrics for evaluating RAG systems. These include both LLM-based and non-LLM-based metrics. The set of multi-turn metrics in the Ragas framework for evaluating RAG systems is very limited. However, there are various approaches for evaluating dialogues in general ~\cite{mehri21}. LLM-based metrics provide the ability for multi-turn-level evaluation without reference solutions. In this case, LLMs assess a response in the context of the dialogue history by considering its semantics.  ~\citet{mehri20} introduced the FED (fine-grained evaluation of dialog) metric, an automated evaluation metric for dialogues that measures eighteen composite dialogue qualities without using reference solutions. Eight of these relate to the turn level and ten to the dialogue level, including properties such as coherence and consistency of the dialogue. 

All the mentioned RAG evaluation approaches rely on static datasets and do not adapt to the RAG system’s actual outputs; instead, they follow a fixed, predefined flow. However, one evaluation framework recently introduced a dynamic evaluation workflow for assessing Conversational AI systems~\cite{levi_intellagent_2025}. This approach employs an LLM agent to actively interact with the system being evaluated.
The evaluation method is built around three main components. First, an event generator creates task-specific events based on predefined policies. These events represent realistic use cases, such as booking or canceling a flight, that the conversational AI is expected to handle. Second, an LLM agent simulates user interactions by engaging in conversation with the system to complete the event-driven task. Third, a conversation critic analyzes the interaction and reports which policies were successfully fulfilled and which were not. This dynamic, task-oriented evaluation provides a more realistic assessment of system performance, as it adapts to the system’s responses and tests its ability to handle interactive, goal-driven conversations.

While significant progress has been made in RAG evaluation, particularly through LLM-based metrics and the introduction of multi-turn datasets, a critical limitation persists. Existing approaches are often not specialized for RAG-specific challenges and the nuances of textual multi-turn dialogue. Instead, they either check only for simple success or rely on static, predefined test data. This static nature prevents the evaluation from dynamically responding to a system's errors or successes, leading to a less realistic conversational flow that fails to capture the true performance of RAG systems in adaptive, real-world scenarios. To overcome these limitations, we introduce our novel evaluation framework: RAG-DIVE.

\section{Evaluation Strategy}
\label{Evalaution_strategy}
\begin{figure*}[ht]
    \centering
    \includegraphics[width=\linewidth]{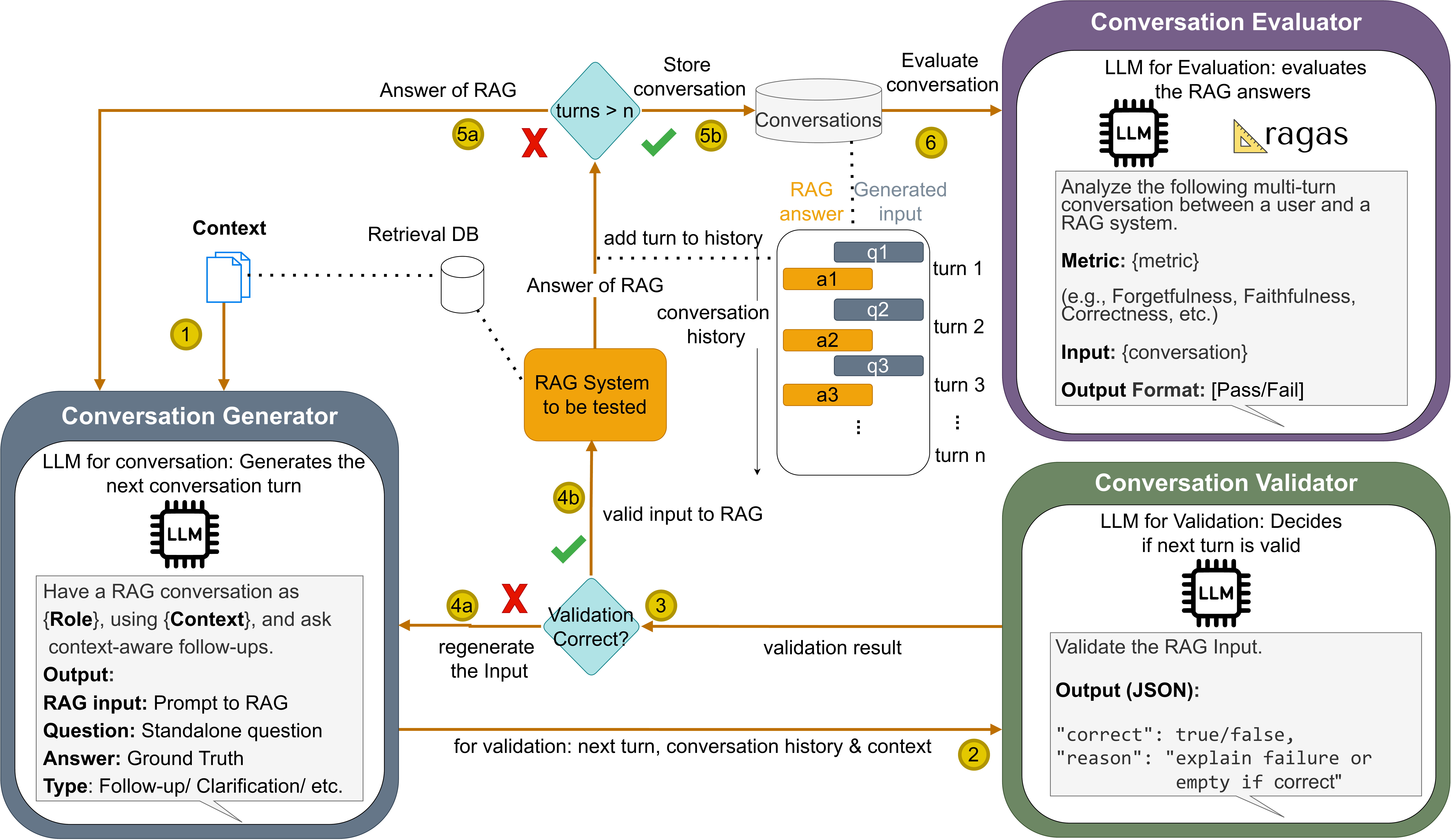}
    \caption{Illustration of RAG-DIVE}
    \label{fig:RAG_DIVE}
    \Description[RAG-DIVE Figure]{Illustration of RAG-DIVE showing the interaction between CG, CV, and CE.}
\end{figure*}
RAG-DIVE automatically evaluates RAG-based multi-turn conversations, dynamically responds to the conversation, and does not require a static QA dataset. This evaluation approach consists of three components:
\begin{enumerate*}[label=(\arabic*)]
    \item \textbf{Conversation Generator (CG)}: Generates questions based on the previous conversation and the context document.
    \item \textbf{Conversation Validator (CV)}: Validates the quality and coherence of the generated questions.
    \item \textbf{Conversation Evaluator (CE)}: Evaluates the RAG model’s responses within the conversation.
\end{enumerate*}
The overall process is illustrated in Figure~\ref{fig:RAG_DIVE}.
\subsection{Conversation Generator} The CG is used to simulate a human interacting with the RAG system. For this purpose, an LLM is employed to simulate a user. It initiates a conversation and, based on the RAG’s responses, continues with a multi-turn dialogue. The CG generates the dialogue based on a provided document. This document is stored within the RAG system, ensuring that the model can retrieve and use its content to answer the questions.

\subsubsection{Input}
First, a \textbf{role} is provided to the CG, defining a realistic user persona for testing the RAG system. This enables the evaluation of system stability across different user types. This design addresses the issue that RAG performance is highly dependent on the user and the way questions are formulated ~\cite{cao_out_2025, percin_investigating_2025}.  Roles are designed to reflect how different users might interact with the system. For example, one role represents a user who asks short and precise questions, while another simulates a user who tends to repeat themselves, overlooks parts of the answer, and sometimes poses repetitive or even factually incorrect questions. These roles are simulated by the CG to test the ability of the RAG to handle diverse conversational behaviors.
Second, the CG is provided with a \textbf{context document} (cf. Figure \ref{fig:RAG_DIVE}, \goldcircled{1}), from which all questions are generated. This document is part of the knowledge base of the RAG system. The CG autonomously decides which questions to ask based on the document, but is restricted to questions that are both related to and answerable from the given content.
For all turns after the initial one, the CG is also provided with the \textbf{RAG’s previous answer} \goldcircled{5a} and generates the next question in a way that is coherent with the prior conversation, maintaining context awareness across turns.

\subsubsection{Output}
Based on the provided information, the LLM is guided to generate four outputs: the RAG input, the stand-alone question, the ground truth answer, and the question type.
\begin{enumerate*}
    \item \textbf{RAG Input}: The direct input sent to the RAG system. This is a question that must be answerable by the RAG, adapted to the assigned role. Except for the first turn, each RAG input must be a continuous question that depends on the previous conversation and is context-aware, rather than fully stand-alone.
    \item \textbf{Stand-alone Question:} A reformulation of the RAG input. In the first turn, it is identical to the RAG input. In later turns, when the RAG input depends on the preceding conversation, the stand-alone version must be rewritten to contain all necessary information to be self-contained while preserving the original meaning. This enables the evaluation of each turn independently, allowing Correctness to be assessed without relying on prior context.
    
    \item \textbf{Ground Truth Answer:} The correct answer to the RAG input, derived directly from the provided document.

    \item \textbf{Question Type:} The classification of the question into one of the categories introduced in ~\cite{sirdeshmukh_multichallenge_2025}- Follow-up, Clarification, Correction, or Comparative.
\end{enumerate*}

This output is sent for validation to the CV \goldcircled{2}, and once validation is complete, the question is forwarded to the RAG system \goldcircled{4b}. The resulting answer is then sent back to the CG as input for the next turn \goldcircled{5a}. If the maximum number of turns is reached, the conversation is stored \goldcircled{5b}, and the generation process begins again from Step \goldcircled{1}.
This process results in a set of conversations generated for each run and each configuration, individually tailored to the provided document and the corresponding RAG answers.

\subsection{Conversation Validator} The CV verifies the output generated by the CG. This step can be seen as a modified form of a chain-of-verification ~\cite{dhuliawala_chain--verification_2023}. To achieve this, an LLM is employed to ensure that all required variables are correctly specified. 
Based on the conversation history and the provided document, the validator decides in a single prompt whether the CG output is valid:
\begin{enumerate*}
    \item \textbf{ The RAG input} must be context-dependent on the preceding conversation and relevant to the provided document. In practice, this means the question does not restate all details explicitly but assumes knowledge of the prior conversation. For example, if the first question is “What is the Eiffel Tower?”, the valid follow-up RAG input would be “When was it built?”. An invalid input would be “When was the Eiffel Tower built?”, since it redundantly restates the subject instead of relying on the conversation context.
    \item \textbf{ The stand-alone question} must be self-contained while preserving the meaning of the RAG input. Using the same example, the correct stand-alone question would be “When was the Eiffel Tower built?”. An invalid formulation would be “When was it built and who designed it?”, since it alters the content and introduces ambiguity.
    \item \textbf{ The ground truth answer} must be validated against the provided document, ensuring that it is factually correct and only includes information supported by the source.
    \item Finally, \textbf{the question type} must match the content of the question.
\end{enumerate*}

The output of the CV is a JSON object containing a Boolean variable indicating whether the output is valid. If the output is not valid, the JSON additionally provides a reason specifying which conditions were not fulfilled and why. This feedback is returned to the CG, which then regenerates the entire turn based on the validator’s response \goldcircled{4a}.

This validation step ensures higher-quality inputs for the RAG system by filtering out unusable outputs from the CG and automatically regenerating new inputs within the same context. To improve this regeneration process, the CV provides explicit feedback, explaining why the CG’s output was incorrect and what aspects require improvement. Based on this feedback, a new turn is generated, resulting in more reliable and coherent conversations.

\subsection{Conversation Evaluator} Finally, the CE takes the list of generated conversations and performs the evaluation \goldcircled{6}. Similar to the previous components, an LLM is employed as a judge, as prior studies have shown that LLM-based evaluation is a reliable and effective method ~\cite{gu_survey_2025, zheng_judging_2023}. Different types of evaluations are conducted, beginning with single-turn evaluation, which compares the standalone question, the RAG’s answer, and the corresponding ground truth answer. This process also considers the retrieved and reference documents to assess how accurately and faithfully the RAG system performs on individual turns. This allows to identify trends, such as whether performance deteriorates over successive turns or remains stable. To measure single-turn performance, we adapt evaluation metrics from Ragas ~\cite{es_ragas_2023}.
Additionally, we conduct multi-turn evaluation, where the entire conversation is assessed. In this setting, we employ the metric of Forgetfulness to measure how well the RAG system retains and leverages information throughout the dialogue, and whether it becomes confused when responding to context-dependent questions. Thanks to the structured output of the CG, it is possible to compute both single-turn and multi-turn metrics. While RAG inputs are typically not self-explanatory or standalone, the inclusion of standalone questions in our framework enables these evaluations.

For our metrics, we adapted the Aspect Critic method, which outputs a Boolean value (0 or 1) for each evaluation, indicating whether the metric is fulfilled, as suggested by Ragas ~\cite{es_ragas_2023}. The final score is the average across all evaluations, resulting in a value between 0 and 1.

Our single-turn metrics are \textbf{Faithfulness}, \textbf{Context Precision}, \textbf{Context Recall}, and \textbf{Correctness}. These metrics are provided by Ragas ~\cite{es_ragas_2023}, demonstrating how RAG-DIVE can easily incorporate existing metrics due to its structured output.
Each single-turn metric requires several inputs: the question, for which we use the stand-alone question generated by the CG. The ground truth answer, also provided by the CG; the reference document, which is the CG’s input document, and the RAG output along with the retrieved documents, which are automatically returned by the RAG system. This setup enables the evaluation of multiple metrics from the Ragas framework.
For the Correctness metric, we implemented a custom evaluation, where each turn is assessed by an LLM to determine whether the answer is correct.

For our multi-turn metrics, we use the whole conversation as input, including both the RAG input and the RAG answer for each turn, organized as a sequential list. We defined two metrics to evaluate performance across turns. \textbf{Forgetfulness} measures whether the RAG system loses information from previous turns, with a score of 0 indicating no Forgetfulness. 
The consistent reproduction of information and coherent conversational flow are two key qualities used to assess multi-turn dialogues~\cite{mehri20}. Building on these evaluation criteria, we introduce \textbf{Context Retention}. Context Retention assesses whether the system successfully retains relevant information from earlier in the conversation, demonstrating understanding and continuity. For this metric, as well as for all other metrics except Forgetfulness, a score of 1 represents the best performance.

The RAG-DIVE framework introduces several key innovations across its components. The CG creates a Role-based fully dynamic evaluation dataset and automates the generation of multi-turn conversations, enabling flexible and context-specific testing of RAG models. The CV ensures high-quality outputs and incorporates a regeneration loop, allowing iterative refinement of the generated conversations. Finally, the CE integrates both single-turn and multi-turn metrics, providing a unified framework capable of evaluating RAG system performance across multiple dimensions. Together, these components make RAG-DIVE a comprehensive and automated solution for dynamic, multi-turn RAG evaluation.

\section{Experimental Setup}
To validate our approach, we conducted experiments on different RAG settings to assess how well our evaluation method captures their performance. For the evaluation setup, we used Gemini-2.0-Flash ~\cite{noauthor_gemini_nodate} as the model for the CG and CE, and GPT-5 Nano ~\cite{openAI_gpt-5_2025} as the LLM for the CV. Both LLMs were used with default generation settings. We employed zero-shot prompting for all prompts. 
In our experiments, we restricted our CG to a single role that generated short and precise questions.
The CV was employed with a single iteration, meaning it checks the output only once, and no further checks are performed in subsequent iterations.
For the evaluation with the CE component, we employed the Ragas framework, incorporating both our proposed metrics and the predefined metrics provided by Ragas~\cite{es_ragas_2023}.
As data for the RAG experiments, we used the ClapNQ dataset with answerable questions ~\cite{rosenthal2024clapnq} and the SQuAD dataset ~\cite{rajpurkar_know_2018} for the industrial usecase. 

For the validation experiments in Section \ref{valdiation_ragdive}, we sampled 100 Wikipedia articles from ClapNQ to serve as unmodified input for the CG. These documents were then used to generate the conversations for evaluation.
Next, we developed a base RAG system and modified RAG systems for evaluation. 
First, we created the Retriever. Here we used the full set of 300 documents from the ClapNQ dataset, which were processed with a RecursiveCharacterTextSplitter into chunks of 4000 tokens with an overlap of 150. The embedding model used was Google Generative AI embedding-001 ~\cite{noauthor_embeddings_nodate}. 
The retriever database was implemented with PGVector ~\cite{noauthor_langchain-ailangchain-postgres_2025}, and search was conducted using similarity search using cosine similarity. 
For the generator component of the RAG, we used for the base system 10 documents as the context list and always provided the complete conversation history to the LLM. These two settings were modified during our experiments. 
The entire RAG pipeline was developed using LangChain ~\cite{noauthor_langchain_nodate}.

To assess the quality of the synthetically generated conversations, we conducted a \textbf{human evaluation} (Section~\ref{human_validation}) by two individuals. For this, we adapted the metrics and definitions from ~\cite{lee_multi-document_2024}. Specifically, we applied the criteria Correctness, answerability, and plausibility at the turn level, and the metrics diversity and coherence at the dialogue level. The correct metric was applied to the generated ground truth answers rather than the RAG responses, as the goal of this evaluation was to assess the performance of the CG and CV.

To make RAG-DIVE a practical evaluation approach, the results must be highly consistent. To demonstrate this \textbf{consistency}, we run in Section \ref{consistencycheck} our evaluation 10 times with 100 conversations of 5 turns and 10 times with 100 conversations of 7 turns. For each run, we calculated all metrics and then reported the mean, standard deviation, the minimal and maximal value to assess stability. We distinguish between single-turn metrics, including Faithfulness, Context Precision, Context Recall and Correctness, and multi-turn metrics, such as Forgetfulness and Context Retention.

To further evaluate our approach, we \textbf{modified} the \textbf{underlying RAG settings} (Section \ref{sensitifity}). Given the limited number of repetitions, we used minimal and maximal values of the baseline RAG scores (cf. table \ref{tab:consistency}, part 5 turns) as an indicator of notable deviation. Values that exceeded this threshold were marked with an asterisk (*).
In our first experiment (cf. Table \ref{tab:rag_modifcation}, part History Modification), we altered the conversation history to assess whether our method can detect Forgetfulness, which should manifest as a higher Forgetfulness score and a lower Context Retention score. Specifically, we tested three conditions:
\begin{enumerate*}
    \item removing the entire conversation history,
    \item  retaining only the first user input as history, to examine whether the CG generates questions that are not solely dependent on the initial query, and
    \item retaining the last three utterances, representing a more realistic and practical approach to history management.
\end{enumerate*}
In another experiment (cf. Table \ref{tab:rag_modifcation}, part Context Modification), we modified the number of context documents available to the RAG, which we expected to negatively affect both Correctness and retriever performance. We tested two configurations: limiting the system to three context documents and restricting it to only one.

To assess the \textbf{practical applicability} of RAG-DIVE within a real-world RAG system, we applied it in Section \ref{industrial_usecase} to evaluate a RAG system provided by an industry partner and compared the results with those obtained using a static-dataset-based evaluation method. For this purpose, we replaced the original RAG data with the SQuAD dataset and used its questions for evaluation. We selected 100 documents for RAG-DIVE and generated five conversation turns per document. For each document, we additionally sampled five questions from SQuAD, resulting in a comparative evaluation dataset of 500 questions.

The core concept for practical usage is establishing a baseline evaluation and iteratively adjusting individual components to enhance performance. Typical examples include 
\begin{enumerate*}
    \item adjusting the chunking strategy or
    \item replacing the underlying models ~\cite{Wang2024-ep}.
\end{enumerate*}
To simulate this process, we established the following baseline configuration: embedding model \textit{multilingual-e5-large-instruct-Q8}, the reranker \textit{bge-reranker-large-q8}, using 5 chunks with 300 tokens per chunk, and the generator model \textit{qwen-4B-q8}. After that, we applied the following scenarios: 
\begin{enumerate*}
    \item {Change Chunking Strategy}. First, the chunk size was increased to 500 tokens to examine whether larger segments improve performance, acknowledging the trade-off of higher token consumption. We also varied the number of retrieved chunks: a single chunk (direct match, enabled by hybrid retrieval with reranking), five chunks (the default setting), and ten chunks (in the large-context configuration).
    
    \item {Change Embedding/Reranking Models}. Initially, we replaced the \textit{multilingual-e5-large-instruct-Q8} embedding model with another high-performing variant, \textit{bge-m3-Q8}. Subsequently, we switched to an English-only embedding model, \textit{bge-small-en-Q4\_K\_M}, which is approximately one-tenth the size of the multilingual model. In addition, the reranker was changed from \textit{bge-reranker-large-q8\_0} to the more performant \textit{jina-reranker-v1-tiny-en-Q8} in conjunction with \textit{bge-small-en-Q4\_K\_M}.
\end{enumerate*}
To indicate notable performance differences, we defined a threshold of 5\% change. Any metric exceeding this threshold is marked with arrows ($\downarrow$, $\uparrow$) to denote a decrease or increase, respectively.

All results and the complete implementation of RAG-DIVE, including code and experimental configurations, are available in our public repository \url{https://github.com/lorenzbrehme/RAG-DIVE.git}.

\section{Validation of the RAG-DIVE Approach}
\label{valdiation_ragdive}
In the following section, we present experiments validating our evaluation approach using a dedicated RAG system. Specifically, we analyzed the conversation flow and CV performance, conducted human assessments of dialogue quality, tested consistency across repeated runs, and evaluated sensitivity to RAG configuration changes.

\subsection{Flow Analysis of Question Interaction}
First, we analyzed the conversation flow by examining how it adapted across different question types.
Five distinct question types were used in our evaluation. The first type, the initial question, always initiates the conversation and is not generated by the CG. The remaining question types are determined by the CG itself. In our 5-turn evaluation setting, the average number of question types per turn was as follows: Follow-Up 83.13\%, Clarification 11.43\%, Comparative 3.95\%, and Correction 1.50\%. These selections were made autonomously by the LLM without any external influence.
Follow-up turns consist of questions that continue from the previous question, maintaining the conversation’s continuity. Clarification questions request additional details from the RAG answer or aim to restate information more clearly. The dynamics of the conversation flow are particularly evident in correction questions. In these cases, when the RAG provides an incorrect or incomplete response, the CG generates a new question based on that output, allowing the system to provide the correct answer.
Example 1 demonstrates the realistic flow of the conversation and shows how the CG identifies and corrects misunderstandings. It indicates human-like behavior and simulates a realistic RAG conversation.
\begin{tcolorbox}[title=Example 1: CG Correction]
\label{fig:example 1}
\textbf{RAG Input:} \\
Who directed the music video for "Be Careful?

\textbf{RAG Answer:} \\
The music video for "Be Careful" was directed by Jora Frantzis.

\textbf{RAG Input:} \\
 When was it released?
 
 \textbf{RAG Answer:} \\
"Be Careful" by Cardi B was released on March 29, 2018. 
[Actual the video release was on May 21, 2018]

\textbf{RAG Input:} \\
 I meant the music video.
\end{tcolorbox}

For comparative questions, the CG incorporates new information seamlessly without referencing the previous context or conversation, further demonstrating a realistic flow of dialogue as provided in Example 2.

\begin{tcolorbox}[title=Example 2: CG Comparison]
\label{fig:example 2}
\textbf{RAG Input:} \\
So, what happens if they can't even get 8 jurors to agree?

\textbf{RAG Answer:} \\
If in Scotland, the jurors cannot get the support of at least 8 jurors for a guilty verdict, it is treated as an acquittal.

\textbf{RAG Input:} \\
 Is that different in civil cases there?
\end{tcolorbox}

\subsection{Relevance of the CV in RAG-DIVE}
To assess the performance of the CV, we logged all instances in which it detected an invalid CG output. 
In our experiments, where we ran 100 conversations of 5 turns each ten times, an average of 25.68 out of 100 outputs did not pass validation.
We manually reviewed these cases and observed several interesting behaviors. 
For this analysis, we manually classified the outputs of the RAG into six distinct categories. The first category was \textit{RAG\_InputNotConnected}, which refers to cases where the RAG input was not related to the question. The second was \textit{notStandalone}, where the standalone question was not self-contained and relied on previous context. The third was \textit{questionNotConsistent}, which captures inconsistencies between the standalone question and the RAG input. The fourth category was \textit{groundTruth}, covering issues with the ground truth answer, such as being incorrect, incomplete, or only partially addressing the question. The fifth was \textit{type}, where the assigned question type was incorrect. Finally, we identified \textit{documentMentioned}, where either the question or the answer explicitly mentioned the document.

In total, we reviewed 109 outputs from one run of 100 conversations, each with 5 turns. In most cases, 49 in total, the data fell into the \textit{groundTruth} category. The distribution across the other categories was as follows: 27 instances of \textit{notStandalone}, 17 of \textit{documentMentioned}, 9 of \textit{RAG\_InputNotConnected}, 4 of \textit{questionNotConsistent} and 2 of \textit{type}. 
In addition, one formatting issue was observed, where the CV states that the CG generated all variables in lowercase. This error 
was not a real error and should not be stated by the CV.

In the following is an example where the CV criticized the ground truth answer. In these cases, the CG misunderstood the context, and the CV correctly flagged and prompted a correction. 

\begin{tcolorbox}[title=CV Example 1: GroundTruth]
\label{fig:CV example 31}
\textbf{Original Output:} \\
The document does not specify other types of goods used in bartering arrangements beyond agricultural produce in exchange for staples and services.

\textbf{Reason of CV:} \\
The provided Answer is factually incorrect: the referenced context indicates that, in addition to agricultural produce, a variety of other goods were bartered (e.g., sugar, coffee and tea, petroleum distillates, textiles like bolts of cloth, needles, and thread, medicines, hardware, and discretionary items like candy or books), and services from professionals were also exchanged via barter. Therefore, it was not limited to agricultural produce alone.

\end{tcolorbox}

The next example is for the \textit{notStandalone} class when the stand-alone question is not truly self-contained or references information from the previous conversation (cf. CV Example 2).

\begin{tcolorbox}[title=CV Example 2: notStandalone]
\label{fig: CV example 2}
\textbf{Original Output:} \\
 Following our discussion about context effects, can you provide an example of how they are specifically used in marketing strategies?\\
\textbf{Reason of CV:} \\
The 'Question' is not standalone. It begins with 'Following our discussion...' which references a prior turn, so it does not rephrase the RAG\_input in a self-contained manner. The Question should be a standalone restatement of the rag\_input without referencing prior discussion or history.
\\

\end{tcolorbox}

\subsection{Human  Validation of Conversation}
\label{human_validation}
\begin{table}[ht]
    \centering
    \caption{Conversation Data Quality (sample size: 200 turns)}
    \begin{tabular}{p{1cm}p{1cm}p{1.2cm}p{1cm}p{1cm}p{1cm}}
    \toprule
        & correct & answerable & plausible & diverse & coherent  \\
         \midrule
        Tester1 & 90.5\% & 87.5\% & 93.5\% & 95\% & 92.5\% \\
        Tester2 & 92.5\% & 87.5\% & 96.5\% & 97.5\% & 95.0\% \\
        \bottomrule
    \end{tabular}
    \label{tab:data_quality}
\end{table}

Table~\ref{tab:data_quality} presents the results of our human validation. In all categories, the scores were above 90\%, except for the answerability score. This metric reached 87.5\%, indicating that 12.5\% of the cases were identified as unanswerable. We observed that in cases where a question was deemed unanswerable, it typically referred to information introduced by the RAG rather than content from the provided document. In such situations, the CG adapted to the RAG’s output and continued asking plausible questions, but occasionally drifted away from the provided document. The corresponding ground truth answer correctly indicated that the question could not be answered with the given document. This shows that the CG can sometimes become distracted and fail to remain strictly within its role of asking questions based only on the provided document. Nevertheless, in these cases, the RAG inputs were still highly plausible and supported a coherent dialogue flow. Overall, in over  90\% of the cases, the answers were correct, and the conversations exhibited a natural, human-like progression with a realistic flow.

\subsection{Consistency Generation Check}
\label{consistencycheck}
To ensure RAG-DIVE’s practicality as an evaluation framework, its results must be \textbf{consistent}, meaning that repeated evaluations under the same conditions produce stable and reliable outcomes across multiple runs. As shown in Table~\ref{tab:consistency}, the standard deviations for both single-turn and multi-turn metrics were low, with a maximum of 0.03 and all other values were below 0.03. Despite differing sample sizes (100 for multi-turn and up to 700 for single-turn experiments), no notable variations were observed. The largest range between minimum and maximum scores was 0.07, averaging 0.04 across metrics. These results confirm that RAG-DIVE delivers stable and repeatable evaluations across runs.
\begin{table}[ht]
    \centering
    \caption{Evaluation Consistency}
    \begin{tabular}{p{2.7cm}p{1cm}p{1cm}p{1cm}p{1cm}}
    \toprule
     \multicolumn{5}{c}{\textbf{Quality of CG} }\\
          \midrule
          metric & mean & std & min & max \\
          \midrule
          
          \multicolumn{5}{c}{\textbf{Quality of CG 5 turns} }\\
          \midrule
         Context Retention & 0.98  &0.01&0.96&1.00 \\
         Forgetfulness & 0.08&0.02&0.06&0.12 \\ 
         Correctness   &0.96&0.01&0.94&0.98 \\
         Faithfulness &  0.94&0.01&0.93&0.96 \\
         Context Precision &  0.76&0.02&0.72&0.78 \\
         Context Recall &  0.93&0.02&0.91&0.96 \\
         \midrule
         \multicolumn{5}{c}{\textbf{Quality of CG 7 turns} }\\
         \midrule
         Context Retention & 0.99  &0.01&0.97&1.00 \\
         Forgetfulness & 0.1&0.03&0.06&0.13 \\ 
         Correctness   &0.96&0.00&0.95&0.97 \\
         Faithfulness &  0.93&0.01&0.91&0.94 \\
         Context Precision &  0.75&0.02&0.72&0.77 \\
         Context Recall &  0.93&0.02&0.89&0.94 \\
         
         \bottomrule
    \end{tabular}
    \label{tab:consistency}
\end{table}

Additionally, we evaluated the \textbf{similarity of conversations} across turns. Thus, we conducted five independent runs using the same set of documents and compared whether the generated questions expressed similar intent and information focus. For example, in one run the CG asked ``Who directed the film?'', while in another it asked ``Who was responsible for directing the movie?'' both differing in phrasing but identical in intent and informational scope. 

We observed that in the first turn, only 39\% of the questions were unique, while the remaining questions were quite similar in meaning (cf. \autoref{tab:similarity}). In fact, in 20\% of the cases, all five conversations began with the exact same question. However, this similarity decreased with each subsequent turn, and by the fifth turn, 99\% of the conversations produced distinct questions. 
This demonstrates the dynamic nature of our approach, as it adapts and reacts based on the RAG’s answers.
\begin{table}[ht]
    \centering
    \caption{Individual Questions per turn (sample size: 200 turns)}
    \begin{tabular}{p{1cm}p{1cm}p{1cm}p{1cm}p{1cm}p{1cm}}
    \toprule
        Average & turn 1 & turn 2 &turn 3 &turn 4 &turn 5 \\
         \midrule
       78.5\% & 39\% & 69.5\% & 88.5\% & 96.5\% & 99\% \\
         \bottomrule
    \end{tabular}
    \label{tab:similarity}
\end{table}

\subsection{Verifying Evaluation Sensitivity}
\label{sensitifity}
\label{sec:ArtificialExample}
\begin{table}[ht]
    \caption{Comparison of RAG modifications with the baseline (complete history, context list size = 10; cf. Table \ref{tab:consistency}, 5 turns).}
    \centering
    \begin{tabular}{p{1.6cm}p{0.8cm}p{0.8cm}p{0.8cm}p{0.8cm}p{0.8cm}p{0.8cm}}
    \toprule
        RAG & 
\rotatebox{90}{Context Retention} & 
\rotatebox{90}{Forgetfulness} & 
\rotatebox{90}{Correctness} &
\rotatebox{90}{Faithfulness} & 
\rotatebox{90}{Context Precision} & 
\rotatebox{90}{Context Recall} \\
    \toprule
        
        Base Min & 0.96 & 0.06 & 0.94 & 0.94 &0.72 &0.91  \\
        Base Max & 1.00 & 0.12 & 0.98 & 0.96 &0.78 &0.96  \\
        \midrule
        \multicolumn{7}{c}{\textbf{History Modification} }\\
        \midrule
       No History  & \textbf{0.73*}& \textbf{0.63*} & \textbf{0.85*} & 0.93& \textbf{0.43*}& \textbf{0.57*}\\
       \midrule
       Initial Turn  & \textbf{0.91*}& \textbf{0.19*}& 0.95& 0.95& 0.76& 0.94\\
       \midrule
        Last 3 Turns & 0.99& 0.06& 0.97& 0.95& 0.77& 0.95 \\
        \midrule
        \multicolumn{7}{c}{\textbf{Context Modification} }\\
        \midrule
        1 Context & 0.99& \textbf{0.16*}& \textbf{0.90*}  &\textbf{0.85*}& \textbf{0.65*}& \textbf{0.60*}\\
        \midrule
        3 Contexts & 0.97& 0.11&0.96& 0.96& 0.78& \textbf{0.88*} \\
       \bottomrule
    \end{tabular}
    \label{tab:rag_modifcation}
      \raggedright
    \footnotesize    
    * Values outside the min–max range of the consistency baseline.
\end{table}
To further verify the sensitivity of our approach, we systematically modified the underlying RAG configurations.
In the first experiment (cf. Table \ref{tab:rag_modifcation}, part History Modification), we \textbf{altered the conversation history}.
When the RAG system operated without access to prior conversation history, the evaluation revealed a clear decrease in Context Retention and a corresponding increase in Forgetfulness. In addition, the retriever performance declined, leading to lower Context Precision and recall, and ultimately to reduced overall Correctness. For the condition where only the initial user query was retained, we observed reduced Context Retention and higher Forgetfulness, but retriever performance and Correctness remained unaffected, as the initial prompt provided sufficient context to retrieve the relevant documents. Finally, the condition using the last three utterances as history did not show any decline in performance. In this realistic modification, the CG generated questions that were both answerable and comprehensible, even without access to the earliest parts of the conversation.
In another experiment (cf. Table \ref{tab:rag_modifcation}, part Context Modification), we \textbf{modified the number of context documents} available to the RAG. In both cases, Context Recall decreased, with the one-document setting performing substantially worse than the three-document setting. A similar trend was observed for Forgetfulness. With three context documents, the other metrics remained largely unaffected, showing only a slight improvement in Context Precision--likely due to the reduced number of irrelevant documents--and in Faithfulness. In contrast, the one-document configuration led to further declines in Context Precision, Faithfulness, and overall Correctness.

\section{RAG-DIVE in an Industrial Setting}
\label{industrial_usecase}
\begin{table*}[ht]
 \renewcommand{\arraystretch}{1.3} 
    \caption{Comparison to the baselines: Using \textit{multilingual-e5-large-instruct-Q8} for embedding and \textit{bge-reranker-large-q8} for reranking, with 5 chunks of 300 tokens; the generator \textit{qwen-4B-q8} remained unchanged.}
    \centering
    \begin{tabular}{
        p{0.4cm} 
        p{4.5cm} 
        p{1.3cm} 
        p{1cm} 
        p{0.7cm} 
         p{0.9cm} 
        p{0.7cm} 
        p{0.9cm} 
        p{0.7cm} 
         p{0.9cm} 
        p{0.7cm} 
        p{0.9cm} 
    }
    \toprule
     & \multirow{2}{*}{\textbf{Category}} &
     \textbf{Context \newline Retention} & 
     \textbf{Forget-fulness} &
      \multicolumn{2}{p{1.6cm}}{\centering\textbf{\newline Correctness}} &
      \multicolumn{2}{p{1.6cm}}{\centering\textbf{Faith-fulness}} &
     \multicolumn{2}{p{1.6cm}}{\centering\textbf{Context \newline Precision}} & 
     \multicolumn{2}{p{1.6cm}}{ \centering\textbf{Context \newline Recall}} \\ 
    \cmidrule(lr){3-3} \cmidrule(lr){4-4} \cmidrule(lr){5-6}\cmidrule(lr){7-8} \cmidrule(lr){9-10} \cmidrule(lr){11-12}
    & & DIVE  & DIVE & DIVE & SQuAD & DIVE & SQuAD & DIVE & SQuAD & DIVE & SQuAD   \\
    \midrule
      \multicolumn{2}{l}{\textbf{Baseline - 5 Chunks, each 300 Tokens}}
       & 1.00   & 0.07 & 0.96 & 0.99 & 0.61  & 0.93 &  0.67& 0.91 & 0.78 & 0.97   \\
    \midrule

    \multirow{3}{*}{%
  \rotatebox[origin=c]{90}{\shortstack{\textbf{Changes}\\\textbf{to Chunks}}}%
}
      & \textbf{Chunk Size 500 Tokens}
      &  0.98    & 0.10   & 0.95  & 1.00& 0.60 & 0.92 & 0.69  & 0.91 & 0.80 & 0.97    \\

      & \textbf{Number of Chunks: 1}
        & 1.00   & 0.09 & 0.93 & 0.95 & 0.44 $\downarrow$ & 0.85 $\downarrow$ &0.57 $\downarrow$  & 0.86 $\downarrow$ & 0.59 $\downarrow$ & 0.85 $\downarrow$    \\
      

      & \textbf{Number of Chunks: 10}
       & 0.98    & 0.04  & 0.98 & 0.99 & 0.60 & 0.94 & 0.63 & 0.87 & 0.80  & 0.98    \\
    \midrule

    \multirow{3}{*}{%
  \rotatebox[origin=c]{90}{\shortstack{\textbf{Changes}\\\textbf{to Modells}}}%
}
      & \textbf{Embedd. \textit{bge-m3-Q8\_0}}
       & 0.99   & 0.08& 0.96 & 0.98  & 0.61 & 0.89 & 0.48 $\downarrow$  & 0.66 $\downarrow$ & 0.79 & 0.95   \\

      & \textbf{Embedd. \textit{bge-small-en-Q4\_K\_M}}
       & 1.00  & 0.08 & 0.97 & 0.98   & 0.58 & 0.93  & 0.63 & 0.90 & 0.75 & 0.96   \\

      & \centering\textbf{+ \textit{jina-reranker-v1-tiny-en-Q8\_0}}
       & 0.87 $\downarrow$     & 0.38 $\uparrow$  & 0.86 $\downarrow$ & 0.92 $\downarrow$  & 0.59  & 0.88 $\downarrow$    & 0.60 $\downarrow$& 0.79 $\downarrow$ & 0.75 & 0.94   \\
    \bottomrule
    \end{tabular}
    \raggedright
    \footnotesize    
    $\uparrow$, $\downarrow$ = large change ($\geq$5\%)
    \label{tab:evaluation_comparasion}
\end{table*}

To verify that RAG-DIVE performs effectively in real-world scenarios, we applied our evaluation approach to an industrial RAG system (details available in the repository) and compared it against a traditional static evaluation method using the same metrics as RAG-DIVE. The results of this comparison are summarized in Table~\ref{tab:evaluation_comparasion}. 
Overall, looking at the single-turn metrics, the multi-turn metrics, Forgetfulness and Context Retention, applied only to RAG-DIVE, as the SQuAD benchmark is limited to single-turn interactions. These metrics indicated performance degradation only when the re-ranker was modified. It can be observed that RAG-DIVE exhibits behavior changes similar to those of the SQuAD benchmark, as shown in \autoref{tab:evaluation_comparasion} in two scenarios. The only difference is, that the RAG-DIVE framework showed lower performance across all metrics compared to the SQuAD benchmark, except for Correctness, where it remained comparable with only a 3\% lower score. This can be attributed to the implementation of the retriever component, which relies solely on the user’s current input to perform the search. In RAG-DIVE, these inputs are context-dependent and often not meaningful in isolation, leading to weaker retrieval quality and thus lower overall performance in the multi-turn setting compared to the single-turn evaluation.

In the first practical case, where we \textbf{modify the chunking strategy}, no positive effect is observed from increasing either the number of tokens per chunk (from 300 to 500) or the number of chunks (from 5 to 10). Therefore, for this factoid single-hop question, the baseline configuration appears to be both appropriate and more efficient. In an extreme scenario, where we assume that one chunk always contains the ground truth, a degradation in performance is observed. For instance, Context Precision decreases in both evaluation approaches: from 0.91 to 0.86 in SQuAD and from 0.67 to 0.57 in RAG-DIVE.
In the second case, where we \textbf{alter the embedding/reranker models}, RAG-DIVE performs similar to SQuAD. Replacing the baseline embedding with another established model, bge-m3-Q8, yields no significant change, except for Context Precision. For Context Precision, although the relevant chunks are retrieved, those containing the ground truth are ranked lower, leading to lower precision. Interestingly, switching to a smaller and faster embedding model, this time limited to English rather than multilingual, results in neither improvement nor deterioration. This suggests that the choice of embedding model exerts limited influence, likely because the reranker (bge-large) effectively mitigates its impact. When both the embedding and the reranker are replaced with smaller models, performance clearly declines. This suggests that while smaller embeddings may still retrieve the relevant chunks among the top-ranked results, the reranker plays a crucial role in maintaining overall accuracy.

\section{Discussion}
Our results demonstrate that RAG-DIVE produces diverse and coherent conversations with a RAG system, yielding over 90\% plausible and correct outputs for evaluation. This advances current evaluation practices by introducing dynamic, multi-turn conversations that adapt in real time to the responses of the tested RAG system, in contrast to static conversation flows that remain fixed and do not react to the system’s answers. In addition, the inclusion of the CV ensures that the outputs of the CG remain correct and valid, thereby improving both the quality and reliability of the generated conversations. Our experiments demonstrate that the approach consistently detects whether modifications in the RAG system lead to performance changes, making it possible to identify both strengths and weaknesses stably and systematically.
Overall, RAG-DIVE effectively reproduces the behavioral patterns observed in the RAG system with the SQuAD dataset, thereby confirming its suitability for real-world applications.
However, our approach introduces the concern that its dynamic nature could lead to inconsistencies in the results. To address this, we conducted a consistency evaluation, which showed that the standard deviation of results did not exceed 0.03. This demonstrates that the approach produces consistent and reliable outcomes. Other LLM-based evaluation methods also exhibit some variance, as they are inherently non-deterministic ~\cite{gu_survey_2025}. This challenge is not unique to evaluation: RAG systems themselves may produce different outputs for the same input due to the non-deterministic nature of their generator components ~\cite{percin_investigating_2025, atil_non-determinism_2025}.

Previous research has demonstrated that LLMs are a valid method for generating multi-turn conversations ~\cite{lee_multi-document_2024}. Our human validation confirmed that, in our setting, the generated questions were generally plausible, and the corresponding answers and stand-alone questions remained correct, coherent, and diverse. Compared to the results reported by ~\citet{lee_multi-document_2024}, our approach achieved higher correctness while maintaining similar levels of answerability, plausibility, and coherence. The main limitation we observed was that in approximately 15\% of the cases, the questions became unanswerable because the CG was distracted by the RAG’s response. In such cases, the ground truth answer correctly indicated that the question could not be answered based on the provided documents. Nevertheless, this issue highlights the need for improvement. Stricter prompting and more rigorous validation by the CV could help mitigate this problem, which should be investigated in future work. Importantly, this limitation did not affect the consistency of the evaluation results, and the approach was still able to reveal both the strengths and weaknesses of the RAG system.


To test the approach with a different LLM model, we replaced the CG with the GPT-5 model in the five-turn setup from Section~\ref{consistencycheck}. The resulting scores were as follows: Forgetfulness 0.01, Context Retention 0.97, Correctness 0.99, Faithfulness 0.97, Context Precision 0.79, and Context Recall 0.96. These results show improvements across several metrics compared to the evaluation using the same RAG system with the Gemini-2.0-Flash model. Additionally, the CV flagged only 54 invalid CG outputs, compared to an average of 128.4 claims for the Gemini-2.0-Flash model.
This reuslts indicate that switching the CG model may not be beneficial, and the results obtained are not directly comparable. While the GPT-5 model led to better RAG performance, the metrics were calculated using the same procedure, suggesting that the evaluation was effectively less challenging. Additionally, the lower number of invalid outputs flagged by the CV indicates that the GPT-5 Nano used as the CV aligns more closely with GPT-5–generated content than with content from the Gemini-2.0-Flash model. This suggests that it may be advantageous to use different models for the CG and the CV to avoid potential bias, as a model may be more permissive with outputs it generated itself and stricter with outputs from other models, which was also investigated in ~\cite{panickssery_llm_2024}. However, this observation requires further research to provide conclusive evidence, and for now, we can only offer it as a suggestion.

Another limitation of our approach concerns the way the next question is generated. In our experiments, we did not observe any decline in answer quality when modifying the history to include only the last three conversation turns. This outcome is explained by the structure of the conversation generation process, which currently produces only simple follow-up turns. A more sophisticated design could allow the CG to reference much earlier turns or even switch topics during the conversation. While such modifications test specific cases, they cannot cover all possible conversation dynamics. Therefore, evaluation results must be interpreted with awareness of how history is implemented. Future work should explore actual RAG usage patterns and align simulated conversation flows accordingly. Moreover, developing diverse CG roles with distinct communication styles could help analyze how different user roles influence the evaluation outcomes.
Another limitation of our approach is that it currently provides only a single document, which prevents the systematic testing of multi-hop conversations. For this purpose, a static evaluation approach has already been developed ~\cite{tang_multihop-rag_2024}. In practice, restricting the context list to a single document should have a more pronounced negative impact on performance, particularly in settings where multi-hop reasoning is required. This limitation becomes more relevant when evaluating RAG systems designed to handle frequent multi-hop queries. Currently, multi-hop behavior occurs in isolated cases, such as when a provided document is split into multiple chunks and a question spans several of them. However, this is not enforced by our method and happens only incidentally. A possible improvement would be to extend the approach to generate multi-hop questions that connect information across multiple chunks deliberately. The challenge lies in ensuring that these chunks are meaningfully related so that the generated questions maintain a natural conversational flow, rather than being based on arbitrarily selected passages.

\section{Threats to Validity}
This section discusses the threats to validity, following the definitions provided by ~\citet{Wohlin2012}.
To ensure \textbf{internal validity}, we aimed to minimize confounding factors. We conducted experiments multiple times to demonstrate both consistency and generalization. Publicly available LLMs were used, none of which were specifically trained for our setting. We also employed publicly available data and a standard RAG system as the evaluation sample. Furthermore, we modified different aspects of the RAG system to demonstrate how changes are reflected in the evaluation results.

To further enhance \textbf{external validity}, we conducted experiments on a second RAG system with a different setup and a different underlying database, demonstrating that the approach works across multiple RAG configurations. This system was provided by an industrial partner, who was not aware of the evaluation approach. Additionally, we repeated the experiments using a different LLM, showing that the method is not limited to a single model. Future work could expand this validation by including additional LLMs, such as smaller models or reasoning-focused models, to explore the approach’s robustness across diverse settings. It should also be noted that some of the system modifications used in our experiments were intentionally extreme (e.g., removing the dialogue history or reducing the context list size to one) to clearly demonstrate the sensitivity of the evaluation approach. While these configurations may not reflect realistic RAG deployments, they serve as controlled test cases. This limits the external validity of our findings, as real-world systems may behave differently under more realistic conditions.
To ensure \textbf{construct validity}, we employed Ragas metrics, a commonly used evaluation approach that provides a valid assessment. Additionally, we introduced our own metrics, which produced similar trends and observations. To further validate the approach, two independent human evaluators assessed the generated conversations, providing an additional layer of human validation and supporting the reliability of our method.

We implemented several measures to ensure \textbf{conclusion validity}. For our consistency experiments, the sample size was relatively small, with only ten samples. To address this limitation, we reported both the minimum and maximum values to illustrate the effects of modifications, rather than relying solely on standard deviations. While this remains a limitation, it allows us to observe trends and visible changes, even if minor, which may not be statistically significant. For other experiments, the sample size was limited to one, which is insufficient for statistical inference. These experiments were intended only to demonstrate that the evaluation approach can detect changes in the RAG system, rather than to prove significant performance differences. Overall, these experiments validate that the approach is capable of identifying modifications and trends in RAG behavior. Additionally, due to the dynamic nature of LLM-based systems, exact reproducibility of the results cannot be guaranteed. This may lead to minor variations in outcomes, even when using the same setup. This limits the repeatability of our experiments and poses a potential threat to the validity of our conclusions.

\section{Conclusion}
This research introduces a new evaluation approach for RAG systems that dynamically adapts to the system’s responses, integrates a validation step to ensure the quality of evaluation questions, and supports adaptive personas to simulate realistic user interactions. The approach consists of three components: The CG is an LLM that interacts with the RAG system under test. The CV validates the outputs of the CG and functions as a chain of verification. The CE evaluates the RAG’s answers and the overall conversation. This approach assesses the RAG in a multi-turn setting and can simulate different user roles. Our experiments demonstrate that this dynamic evaluation method produces consistent and robust results across different RAG configurations and multiple iterations of the same setup. Manual review of the generated conversations confirmed that the outputs are plausible and exhibit realistic conversational flow, supporting their use for evaluating RAG performance.
Future work could extend this approach to evaluate multi-hop settings by leveraging multiple interconnected documents as the context base. Additionally, research could explore alternative structures for conversation history and investigate realistic RAG user behavior to simulate diverse conversational flows. Further studies could focus on how different user roles influence the evaluation workflow and identify which roles are most relevant for each RAG system. These investigations could be conducted in industrial settings with real-world use cases, providing feedback on how the evaluation contributes to improving RAG performance. 
Overall, this approach represents a novel method for automating RAG system evaluation by employing LLMs to simulate and assess multi-turn interactions.

\begin{acks}
This work was partially funded by FFG (Austria, Grant 921318) and BMFTR (Germany, Grant 16IS24069H) within the GENIUS project.
\end{acks}

\bibliographystyle{ACM-Reference-Format}
\bibliography{references}
\end{document}